\documentclass[11pt]{article}
\usepackage{epsfig}
\usepackage{amsfonts}

\def\laq{~\raise 0.4ex\hbox{$<$}\kern -0.8em\lower 0.62
ex\hbox{$\sim$}~}
\def\gaq{~\raise 0.4ex\hbox{$>$}\kern -0.7em\lower 0.62
ex\hbox{$\sim$}~}

\def\beq{\begin{equation}}
\def\eeq{\end{equation}}
\def\bea{\begin{eqnarray}}
\def\eea{\end{eqnarray}}

\def \ra {\rightarrow}

\def \la {\lambda}
\def \ls {\lambda_{\rm s}}

\def \Mp {M_{\rm P}}

\def \La {\Lambda}

\def \a {\alpha}
\def \ap {\alpha^{\prime}}

\def \Ga {\Gamma}

\def \sg {\sigma}

\def \r {\rho}

\def \noi {\noindent}

\begin{document}
\begin{titlepage}

\begin{flushright}
BA-TH/811-26\\
\end{flushright}

\vspace{0.2 cm}

\begin{center}

\LARGE{Cosmological horizons in regular bouncing backgrounds}

\vspace{1cm}

\normalsize
{M. Gasperini}

\bigskip
\normalsize

{\sl Dipartimento di Fisica,
Universit\`a di Bari, \\
Via G. Amendola 173, 70126 Bari, Italy\\
and\\
Istituto Nazionale di Fisica Nucleare, Sezione di Bari, Bari, Italy \\
\vspace{0.3cm}
E-mail: {\tt gasperini@ba.infn.it}

\vspace{0.3cm}
Submission date: {\rm 20 March 2026}}

\vspace{1cm}

\begin{abstract}
\noi
It is often stated that a phase of standard, decelerated cosmological expansion is characterised by the absence of global event horizons, while a phase of accelerated expansion is associated with the absence of particle horizons. This is not necessarily true because such horizons, being non-local properties of the spacetime geometry, depend on the full (past and future) history of the given cosmological background. We provide examples of various different scenarios for the case in which the final asymptotic phase of standard expansion and decreasing curvature is connected, through a regular bounce, with an initial (and possibly infinitely extended in time) regime of growing curvature. 
\end{abstract}
\end{center}

\vspace{0 mm}
\begin{center}
---------------------------------------------\\
\vspace {5 mm}
Essay written for the {\em 2026 
Awards for Essays on Gravitation,}\\
{  (Gravity Research Foundation, Wellesley Hills, MA, 02481-0004)}\\
 and awarded with {\em ``Honorable Mention"}\\

\vspace {5 mm}
To appear in: {\bf International Journal of Modern Physics D (2026)}
\end{center}
\end{titlepage}

\newpage
\parskip 0.2cm

It is well known that there are at least three possible (and physically distinct) types of non-trivial surfaces, conventionally called ``horizons", in cosmological spacetime geometries:  the Hubble horizon, the particle horizon and the event horizon (see e.g. \cite{1}). 

The first one is directly associated, at any given time $t$, with the $2$-sphere centered on the observer and of radius $R_H(t)$ given by the inverse of the local Hubble parameter, i.e. $R_H(t)= |H(t)|^{-1}$; the second and the third ones are determined instead by the ``time of flight'' of light signals emitted by distant sources and traveling towards the observer along null radial geodesics (we are assuming, for simplicity, rotational symmetry along the observer world-line). 

The particle horizon, in particular, is a surface of proper radius $R_P(t)$ separating the portion of space that is causally connected with the observer at the time $t$ -- i.e. that includes all points already ``visible'' to the observer at the given time $t$ -- from the more distant regions of space, whose signals will (possibly) be received later, at times $t_r>t$. 
The event horizon, of  proper radius $R_E(t)$, separates instead the points of space whose signals emitted at the time $t$ will eventually be received in the future by the observer  placed at the origin, from those (more distant) points whose signals will never reach the origin -- hence, will never be observed -- even extending up to infinity (or to its maximal allowed value) the time parameter of the given spacetime-manifold. 

Unlike the Hubble radius $R_H$, the radius of the particle and of the event horizons are thus {\em non-local} quantities, depending, respectively, on the details of the past and future cosmological history. Considering in particular cosmological geometries described by a spatially flat Friedmann-Lemaitre-Robertson-Walker (FLRW) metric, with scale factor $a(t)$, the radius $R_P(t)$ and $R_E(t)$ are given (in cosmic time $t$) by
\beq
R_P(t)= a(t) \int_{t_{\rm min}}^t {dt'\over a(t')}, ~~~~~~~~~~~~~~~~
R_E(t)= a(t) \int_t^{t_{\rm max}} {dt'\over a(t')},
\label{1}
\eeq
where $t_{\rm min}$ and $t_{\rm max}$ correspond, respectively, to the maximal allowed extension of the cosmic-time coordinate towards the past and towards the future for the spacetime manifold that we are considering.  

In the context of the standard cosmological scenario, where the Universe emerges from an initial Big Bang singularity and evolves towards lower and lower curvature states with a monotonically expanding power-law scale factor, the Hubble radius tends to grow linearly in cosmic time, starting from the initial value $R_H=0$. In such a context one finds that a phase of accelerated (inflationary) evolution has no particle horizon, but it is characterised by an event horizon of finite (possibly time-dependent) radius $R_E(t)$; a phase of decelerated evolution, instead, has no event horizon, but is characterised by a  particle horizon of finite radius $R_P(t)$. In that case the proper sizes of the particle and event horizons, when they exist, tend to grow in time proportionally to the Hubble radius $R_H$.
  
Given the above, well known results, a question which may naturally arise is  the following:  what happens to such cosmological horizons in (more exotic)  bouncing scenarios, where the Universe reaches the final regime of standard expansion and decreasing curvature not emerging from a singularity, but smoothly evolving from a regular and infinitely extended pre-bouncing evolution at growing curvature? 
In that case the size of the non-local causal horizons (controlled by $R_P$ and $R_E$), even if finite, might behave very differently from that of the Hubble radius $R_H$. The aim of this paper is to give a concise illustration of the different scenarios possibly existing for the behaviour of the non-local horizons, depending on the different kinematic properties of the cosmic geometry in the pre-bouncing and post-bouncing regimes. 

A first interesting possibility is the case of regular bouncing backgrounds which describe asymptotically a final decelerated expansion (up to $t=+\infty$), and which are characterised by the absence not only of the event horizon but also of the particle horizon (differently from the standard cosmological case). 

Examples of this type can be found within the class of $T$-duality invariant solutions of the string cosmology equations including an appropriate non-local dilaton potential \cite{2}, like, for instance, the following particular exact solution \cite{3}:
\beq
a(t)= \left[t+ (1+t^2)^{1/2} \right]^{1/ \sqrt d}, ~~~~~~
H(t)= \left[ d (1+t^2)\right]^{-1/2}, ~~~~~~ - \infty \leq t \leq + \infty
\label{2}
\eeq
($d$ is the number of spatial isotropic dimensions, and $t$ is given in units of the string length parameter $\ls \equiv \sqrt{\ap}$). Other examples of this type, even without introducing a dilaton potential, can be obtained by considering $O(d,d)$-symmetric string models including higher curvature corrections to all orders in the $\ap$ expansion \cite{4}, like, for instance, the bouncing model described by the following particular exact solution \cite{5,6}:
\beq
a(t)= e^{-{2t\over \sqrt{1+dt^2}}}\left[-\sqrt d\, t+ \sqrt{1+dt^2}\right]^{-{1\over  \sqrt d}}, ~~~~~
H(t)={dt^2-1\over (dt^2+1)^{3/2}}, ~~~~~ - \infty \leq t \leq + \infty.
\label{3}
\eeq

In both cases the solutions  of Eqs. (\ref{2}) and (\ref{3}) describe a regular bounce from growing to decreasing curvature, with a bell-like shape of the (asymptotically positive) Hubble parameter\footnote{Actually, for the solution (\ref{3}), the Hubble parameter $H(t)$ has not a maximum but a local negative minimum at the bounce time $t=0$.}, and with a scale factor evolving from initial accelerated expansion to final decelerated expansion,
\bea
&&
\!\!\!\!\!\!\!\!\!\!\!\!
t \ra - \infty: 
a(t) \sim (-t)^{-1/ \sqrt d}, ~ \dot a >0, ~ \ddot a >0; 
~~~~~~~ t \ra + \infty: 
a(t) \sim (+t)^{1/ \sqrt d}, ~ \dot a >0, ~\ddot a <0
\nonumber \\ &&
\label{4}
\eea
(a dot denotes differentiation with respect to cosmic time $t$). 
 In both cases, the presence of a real-valued and non-singular geometry in the whole temporal range of the solutions is due to the dynamical effects of the string-modified cosmological equations, containing non-local and/or higher-derivative contributions. 
One can then easily check that, in both cases, the Hubble radius $R_H(t)= |a/\dot a |$ is always locally finite, while  the horizon integrals of Eq. (\ref{1}) for $R_P$ and $R_E$ are both locally diverging at any time $t$, given the limiting values $t_{\rm min}=-\infty$ and $t_{\rm max}=+\infty$. Hence we find, for these backgrounds, neither particle nor event horizons. 

Exactly the same result can be obtained if we include the contribution of fluid matter sources in the string cosmology equations, and we look for regular, self-dual bouncing solutions connecting a final phase of radiation-dominated expansion (with equation of state $p/\rho=1/d$, $a \sim t^{2/(d+1)}$ at $t \ra +\infty$) to an initial pre-bouncing phase dominated by a gas of ``frozen" strings (with $p/\rho=-1/d$, $a \sim (-t)^{-2/(d+1)}$ at $t \ra -\infty$), as explicitly discussed in \cite{7} (see also \cite{8} for other bouncing solutions with similar properties). For all such solutions one finds that the integrals of Eq. (\ref{1}) are both diverging, implying the total absence of particle and event horizons. 

In spite of the examples given above, however, it would be wrong to conclude that all self-dual bouncing backgrounds, just because of their intrinsic duality symmetries, are always characterised by the absence of global event horizons. Let us give indeed an example of self-dual solutions of the string cosmology equations which are defined on a regular spacetime manifold spanned by a time coordinate extending from $-\infty$ to $+\infty$, which describe accelerated expansion and which, like in the case of standard cosmology, have no particle horizon but have a finite, time-dependent, event horizon.

\begin{figure}[t]
\centering
\includegraphics[height=4.5cm]{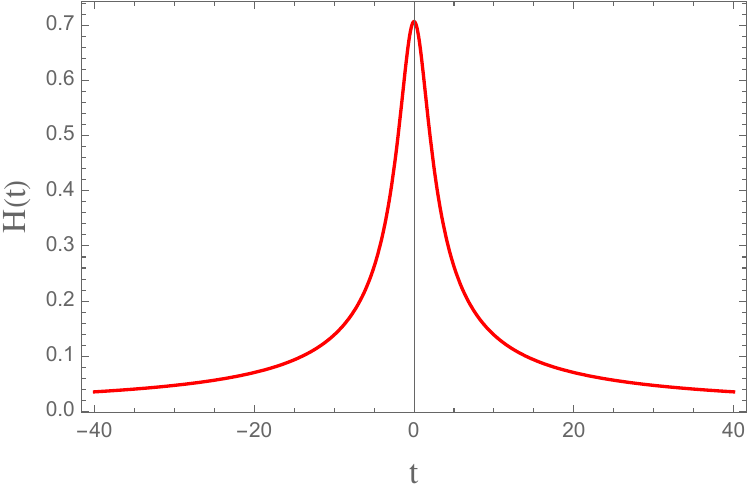} ~~~~
\includegraphics[height=4.5cm]{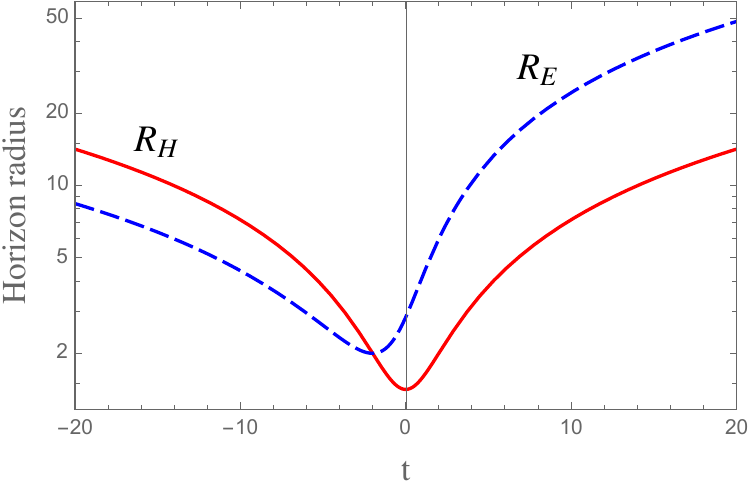}
\caption{Left: the Hubble parameter $H$ for the bouncing solution (\ref{5}). Right: the corresponding radius of the Hubble horizon $R_H$ (red curve) and of the event horizon $R_E$ (blue dashed curve), from Eqs. (\ref{5}), (\ref{6}).}
\label{fig1}
\end{figure}

A class of cosmological backgrounds of this type can be obtained by including string-loop corrections of arbitrary order in the non-local dilaton potential, as shown in \cite{2}. The model depends on three parameters: the loop-order $n$, a constant mass parameter $m$ and a positive dimensionless coefficient $\a$. By limiting here to the case of first-order (one-loop) corrections, a particular exact solution of the string cosmology equations describing a regular bounce can then be obtained by setting $\a=1/2$, $m=1$ (in string units), and can be written as:
\beq
a(t) = \exp\left[ \sqrt 2 {\rm \,Arcsinh}\left(t\over 2\right)\right], ~~~~~~~
H(t)= \left(2\over 4+ t^2\right)^{1/2}, ~~~~~~
 - \infty \leq t \leq + \infty.
\label{5}
\eeq
This solution describes a regular bounce connecting an initial phase of accelerated expansion and growing curvature, with $\dot a >0$, $\ddot a > 0$, $\dot H >0$, to a final, post bouncing phase of accelerated expansion and decreasing curvature, $\dot a >0$, $\ddot a > 0$, $\dot H <0$. There is no particle horizon, as the integral (\ref{1}) for $R_P$ diverges, but there is a finite (time-dependent) event horizon given (from Eqs. (\ref{1}) and (\ref{5}))  by:
\beq
R_E(t)= t + \left( 8+ 2 t^2 \right)^{1/2}.
\label{6}
\eeq
The time behaviour of this event horizon, compared with the Hubble horizon $R_H= |\dot a /a|^{-1}$, is illustrated (on a logarithmic scale) in Fig. \ref{fig1}. 

It may be appropriate to stress, at this point, that to obtain regular bouncing solutions it is not at all crucial to refer to self-dual string cosmology models (as previously done); also, it is not essential to have a growing-curvature, pre-bouncing phase describing asymptotic expansion ($H>0$), because the initial evolution at growing curvature may also correspond to a phase of asymptotic contraction ($H<0$). In this second case, in particular, it turns out that there is the interesting possibility of regular, consistent bouncing scenarios where the particle horizon and the event horizon are both finite and well defined. What is always crucial for a regular bounce, however, is the use of generalised models of gravity in order to evade the standard singularity theorems applying to the Einstein cosmological equations.

Let us present here, in this respect, a simple model of bounce which connects the final regime of standard $\La$CDM expansion to an initial phase of decelerated contraction, and which is based on the cosmological equations of a modified theory of massive gravity \cite{9}. 
Such a theory is different from the standard scenario of massive gravitons because it contains only two propagating  gravitational degrees of freedom, without any additional polarisation mode. Also, the graviton mass term does not contribute to the evolution of the background geometry, as it only appears into the equations for tensor perturbations \cite{9}.
With an appropriate choice of its parameters the model is singularity-free and ghost-free in spite of the presence of an effective graviton mass term, needed to prevent the occurrence of Belinski-Khalatnikov-Lifshitz (BKL) instabilities \cite{10} and to produce a realistic suppression of the tensor-to-scalar ratio of the  perturbations amplified by the initial contracting phase. 

The model includes dust fluid sources with energy density $\rho$, and is described by the following, non-standard cosmological equations (in Planck 
units $ \Mp^2=(8 \pi G)^{-1}=1$, and in $d=3$ spatial dimensions): 
 \beq
 3H^2=\r+\La_1, ~~~~~~~~ 
 2 \dot H=-\r+\La_2,   ~~~~~~~~ 
 \dot \r +3 H \r=0, ~~~~~~~~
 \dot \La_1=3 H\La_2,  
 \label{7}
 \eeq
where  $\La_1$ and  $\La_2$ are time-dependent functions appearing in the effective action of the generalised model of massive gravity \cite{9}. 
These parameters are unrelated to the effective graviton mass and fully independent of its particular  value \cite{9}. 
The function $\La_2(t)$ can be chosen freely while $\La_1(t)$, for consistency, must evolve according to the last of the above equations. In particular, as  shown in \cite{9}, a time-dependent violation of the Null Energy Condition,  and a corresponding phase of regular bouncing, can be efficiently implemented by choosing a function  $\La_2(t)$ of the form
\beq
\La_2(t)= \La_{20} \exp(-t^2/\sg^2),
 \label{8}
\eeq
where $\La_{20}$ and $\sg$ are positive constant parameters.
By adopting appropriate values of the parameters $\sg$, $\La_{20}$ and of the initial conditions we can then numerically integrate the system of Eqs. (\ref{7}), and we obtain solutions describing regular backgrounds with the following geometric properties\footnote{For our example we have used the same numerical  values of the parameters and of the initial conditions at a time $t_i$ as in \cite{9}, namely: $\La_{20}=10^{-4} \Mp^4$, $ \sg= 18.441\, \Mp^{-1}$, 
$a(t_i)=1$, $H(t_i) =-2.758\times 10^{-4}\Mp$,    
$\r(t_i)=2.282 \times 10^{-7} \Mp^4$. However we have changed, for graphical reasons, the initial time, by  choosing $t_i=-16\, \Mp^{-1}$ instead of $t_i=  -1.6 \times 10^3\, \Mp^{-1}$}. 

The effective cosmological constant $\La_1$ tends asymptotically to be positive and constant, and to dominate over $\rho$, both at $t \ra - \infty$ and $t \ra + \infty$ (the constant asymptotic limit of $\La_1$ is however different in the two limits, as the solution is not time-reflection symmetric). The Hubble parameter $H(t)$ tends asymptotically to a constant, positive at  $t \ra + \infty$ and negative at  $t \ra - \infty$. The acceleration of the scale factor, $\ddot a/a$, tends asymptotically to a positive constant both at  $t \ra + \infty$ and at  $t \ra - \infty$. Hence, the resulting background geometry smoothly evolves from {\em decelerated contraction} to {\em accelerated expansion}, the horizons integrals (\ref{1}) are both converging in the limits  $t_{\rm min}=-\infty$ and $t_{\rm max}=+\infty$, and we have a spacetime manifolds with {\em both particle horizon and event horizon}. Their behaviour, compared with that of the Hubble horizon $R_H$, is illustrated (on a logarithmic scale) in Fig. {\ref{fig2}. 

\begin{figure}[h]
\centering
\includegraphics[height=4.5cm]{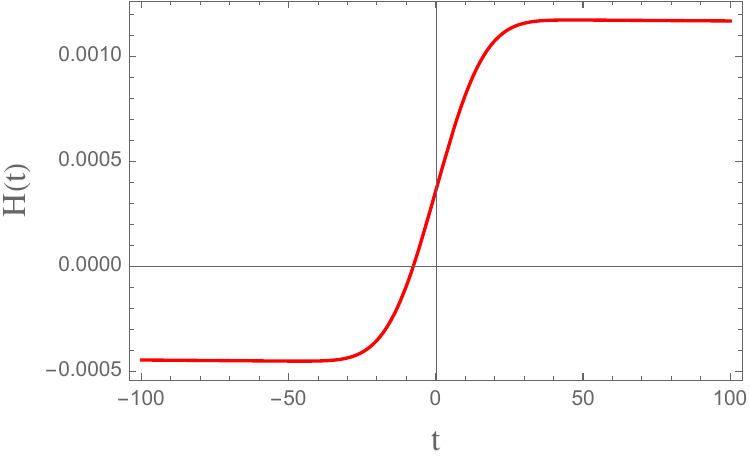} ~~~~
\includegraphics[height=4.5cm]{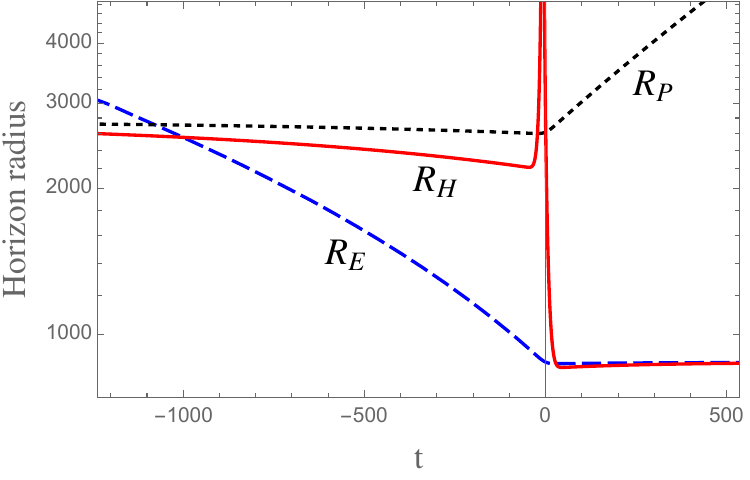}
\caption{Left: the Hubble parameter $H$ for a numerical solution of Eqs. (\ref{7}), (\ref{8}), obtained with the same initial  conditions used in \cite{9}. The background geometry describes a smooth evolution from decelerated contraction to accelerated expansion.  
Right: the associated behaviour of the Hubble horizon $R_H$ (red curve), of the event horizon $R_E$ (blue dashed curve) and of the particle horizon $R_P$ (black dotted curve). The numerical values of the parameters used for the plots are given in Footnote 2.}
\label{fig2}
\end{figure}

Let us complete our discussion by presenting other classes of regular bouncing backgrounds which describe a smooth evolution from decelerated contraction to accelerated expansion and which, like the previous example, are characterised by the presence of both the particle and event horizon. Backgrounds of this type can be obtained in the context of generalised gravitational theories like $f(R)$-gravity or the so-called scalar-tensor models of ``quintom gravity", typically including higher derivatives and non-minimal gravitational couplings. We shall concentrate on two particular examples. 

The first one is described by the following background geometry 
\beq
a(t)= 1+ t^q, ~~~~~~~~~~~~ H(t)= {q\over t(1+t^{-q})}, ~~~~~~~~~~~~ 
 - \infty \leq t \leq + \infty 
\label{9}
\eeq 
(where $q=2n$ with $n$ integer, and we are using Planck units), which can be obtained as a particular exact solution of the cosmological equations in $d=3$ spatial dimensions for a simple model of quadratic gravity \cite{11}. The second example, again in $d=3$ and in Planck units, is given by
\beq
a(t)= \left[ t^2 + {s\over 1-r}\right]^{1/3(1-r)}, ~~~~~~~ H(t)= {2t\over 3\left[s+(1-r)t^2\right]}, ~~~~~~~ 
 - \infty \leq t \leq + \infty
\label{10}
\eeq 
(where $r<1$ is a dimensionless constant and $s>0$ a constant parameter with squared-length dimensions), and can be obtained as a particular exact solution of a quintom model \cite{12} with matter sources satisfying the effective equation of state $p/\rho \equiv w(t) = -r -s/t^2$. Note that for $r=2/3$ and $s=1/3$ one exactly recovers the other example of Eq. (\ref{9}) with $q=2$. 

It can be easily checked that the background geometries (\ref{9}) and (\ref{10}) are both describing regular manifolds evolving from an initial asymptotic decelerated contraction ($\dot a<0$, $\ddot a >0$) at $t \ra -\infty$ to a final asymptotic accelerated expansion  ($\dot a>0$, $\ddot a >0$) at $t \ra+\infty$. In such a context we may expect that both geometries be characterised by the presence of finite particle horizons and event horizons, since the integrals of Eq. (\ref{1}) are both convergent. The explicit results for the horizons of the two  solutions given above are analytically different but have  a very similar qualitative behaviour, so we will report here only a graphic illustration for the solution (\ref{10}), and for the particular case $s=1$, $r=1/2$. In that case one finds
\bea
&&
R_P(t)= \left(1+{t^2\over 2}\right)^{2/3} \left[ \sqrt{8 \pi}\, {\Ga(7/6)\over \Ga(5/3)}+  \,_2{\cal F}_1\left({1\over 2}, {2\over 3}, {3\over 2}; -{t^2\over 2}\right) t \right],
\nonumber \\ &&
R_E(t)= \left(1+{t^2\over 2}\right)^{2/3} \left[ \sqrt{8 \pi}\, {\Ga(7/6)\over \Ga(5/3)}-  \,_2{\cal F}_1\left({1\over 2}, {2\over 3}, {3\over 2}; -{t^2\over 2}\right) t \right],
\label{11}
\eea
where $\Ga(x)$ is the Euler gamma function and $ \,_2{\cal F}_1\left(a, b, c; x\right)$ is the Gauss hypergeometric function. The time evolution of the above expressions for $R_P$, $R_E$, compared with that of the Hubble radius $R_H$, is illustrated (on a logarithmic scale) in Fig. \ref{fig3}. 

It may be interesting to note that for $t>0$ the particle horizon is always ``larger" that the event horizon, $R_P>R_E$. This is to be expected since at a time $t>0$, in a time-symmetric manifold, any given virtual observer is always ``closer" to the future edge of the spacetime manifold than to the past edge: hence, at a time $t>0$, there has been more time available to receive the distant signals emitted in the past, and there is less time available to wait for the distant signals emitted now, to be received in the future. The opposite is true for $t<0$, where we always find $R_P<R_E$ as illustrated in Fig. \ref{fig3}. It is important to stress that this crucial physical property, i.e. $R_P>R_E$ for $t \ra+\infty$ and $R_P<R_E$ for $t \ra-\infty$, is also asymptotically satisfied by the solutions of the ``matter-bounce" scenario of Eq. (\ref{7}) (see Fig. \ref{fig2}, right plot), even if, in that case, the background evolution does not satisfy the property of time-reflection symmetry. 

\begin{figure}[t]
\centering
\includegraphics[height=4.5cm]{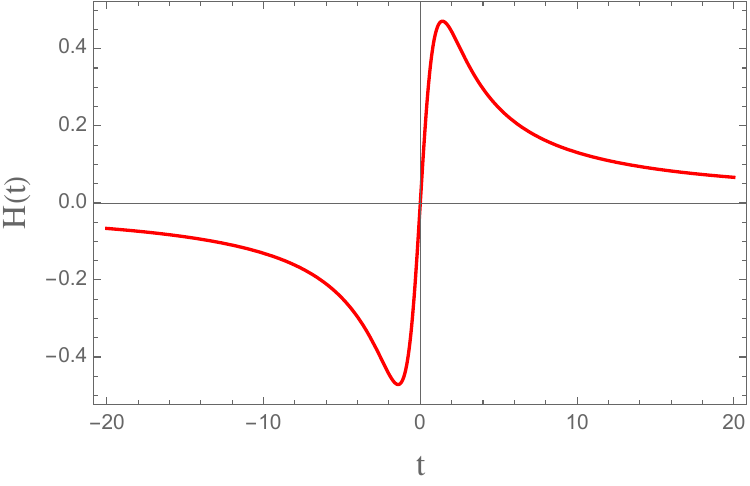} ~~~~
\includegraphics[height=4.5cm]{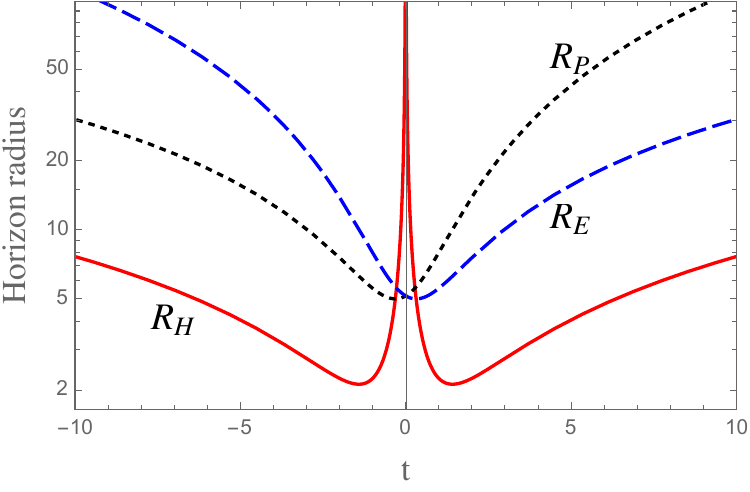}
\caption{Left: the Hubble parameter $H$ for the bouncing solution (\ref{10}), with $s=1$ and $r=1/2$. Right: the associated behaviour of the Hubble horizon $R_H$ (red curve), of the event horizon $R_E$ (blue dashed curve) and of the particle horizon $R_P$ (black dotted curve).}
\label{fig3}
\end{figure}

In conclusion, a few comments are in order. The first is that, in the context of bouncing cosmologies,  the time behaviour of $R_P$ and $R_E$ may be very different from that of the Hubble radius $R_H$, unlike what happens in a standard cosmological context. Also, differently from the standard scenario, the particle horizon may be always existing, with a finite radius, in spite of the simultaneous presence of the event horizon. This provides a counter-example to possible solutions of the horizon problem suggested for bouncing cosmologies \cite{13} and based on the total absence of particle horizons\footnote{See also \cite{14} for a discussion of the horizon problem within a phenomenological model of bouncing.}. 

It should be stressed, finally, that the possible ``unconventional" behaviour of the particle and event horizon presented here have no direct 
consequences at all for the amplification of the quantum metric fluctuations, whose evolution is governed by a different {\em local} horizon $R_\la(t)$, defined by the effective potential $V= a''/a$ appearing in the equation for the Fourier modes of the canonical perturbation variables \cite{15} (a prime denotes differentiation with respect to the conformal time $\eta$, defined by $dt/d\eta=a$). Such an effective horizon is conceptually different, both in principle and in practice, from the other horizons. Its proper radius $R_\la(t)$, controlling the so-called ``horizon crossing" epoch of a Fourier mode of proper wavelength $\la$ (namely, the epoch marking the end of the oscillating regime and the beginning of the phase of parametric amplification) is fixed (in the linear approximation) by the condition $a''/a= k^2 \equiv (a/\la)^2$, and is thus given by
\beq
R_\la(t)= a \left|a''\over a \right|^{-1/2} = \left| \dot H + 2 H^2 \right|^{-1/2}.
\label{12}
\eeq

\begin{figure}[t]
\centering
\includegraphics[height=4.5cm]{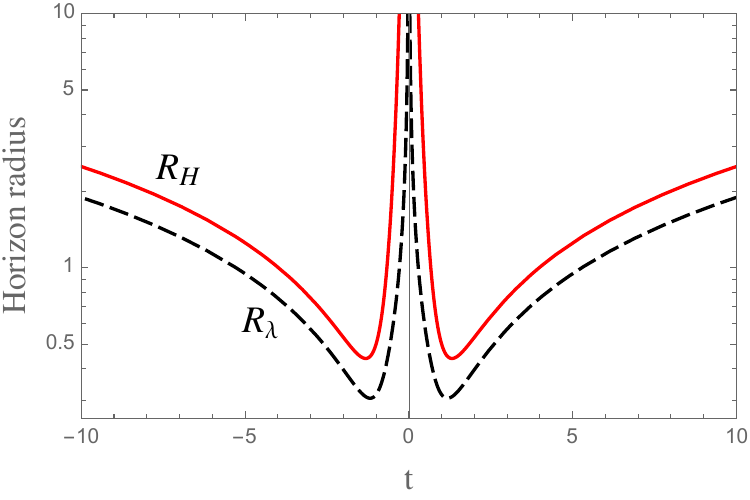} ~~~~
\includegraphics[height=4.5cm]{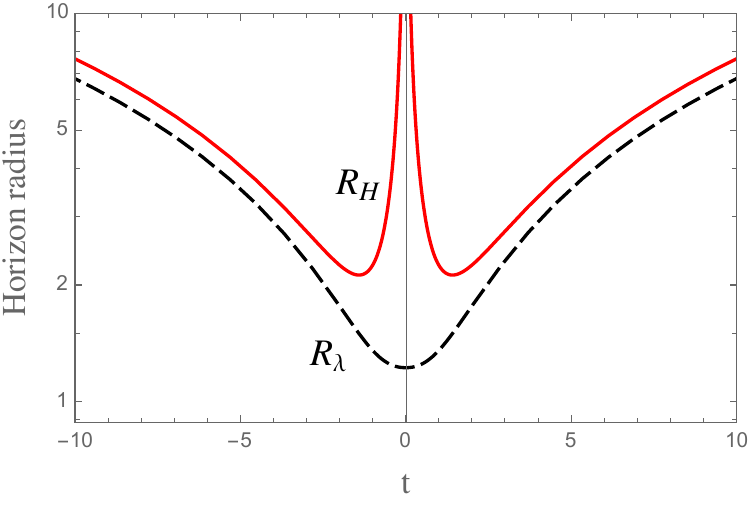}
\caption{Left: the Hubble radius $R_H$ (red curve)  and the local perturbation radius $R_\la$  (black dashed curve) for the bouncing solution (\ref{9}) with $q=4$. Right: the same 
for the bouncing solution (\ref{10}) with $s=1$ and $r=1/2$. }
\label{fig4}
\end{figure}

In spite of these formal and conceptual differences   
it turns out that asymptotically, for a general bouncing scenario, the horizon radius $R_\la$ tends to follow very closely the time behaviour of the standard Hubble horizon $R_H$, with all possible differences localised just around the bouncing region. This can be easily checked by comparing the solutions for $R_\la$ and $R_H$ in all models of bounces presented in this paper: see e.g. Fig. \ref{fig4}, where we have plotted $R_\la$ for the particular bouncing solution (\ref{9}) with $q=4$ (left plot), and for the solution (\ref{10}) with $s=1$, $r=1/2$ (right plot). As a consequence, the kinematic details and/or the intrinsic symmetries of the various models of bouncing can leave peculiar, model-dependent imprints on the primordial spectrum of metric perturbations \cite{16,17} at most on the high-frequency sector of the spectrum (i.e. for those modes crossing the horizon just at the bouncing epoch),  
 quite independently from 
the overall behaviour of $R_P$ and $R_E$.



\section*{Acknowledgements}
It is a pleasure to thank Massimo Giovannini, Jnan Maharana and Gabriele Veneziano for  a long and fruitful collaboration on the general aspects of regular bouncing manifolds satisfying the string cosmology equations and implementing models of the Pre-Big Bang scenario.

\end{document}